\documentclass[10pt]{article}

\usepackage{amsmath}
\usepackage{amsfonts}
\usepackage{graphicx}
\usepackage{graphics}
\usepackage{theorem}
\usepackage{color}

\usepackage{amsxtra}
\usepackage{amstext}
\usepackage{amssymb}
\usepackage{latexsym}

\usepackage[square]{natbib}
\usepackage{graphicx}
\usepackage{dcolumn}
\usepackage{bm}

\makeatletter
\newcommand{\cn}{\mathop{\operator@font cn}}
\makeatother
\makeatletter
\newcommand{\sn}{\mathop{\operator@font sn}}
\makeatother
\makeatletter
\newcommand{\dn}{\mathop{\operator@font dn}}
\makeatother

\newcommand{\Saito} {Sait\^{o}}

\begin{document}


\nocite{*}

\title{Stability of Simple Periodic Orbits and Chaos in a Fermi -- Pasta -- Ulam Lattice}
\author{Chris ANTONOPOULOS$^{\;1}$ and Tassos BOUNTIS$^{\;1}$}
\date{\today}
\maketitle
\begin{center}
{\it $^1\;$Department of Mathematics\\and\\Center for Research and
Applications of Nonlinear Systems (CRANS),\\University of
Patras,\\ GR--$26500$, Rio, Patras, Greece}\\
\end{center}

\begin{abstract}
We investigate the connection between local and global dynamics in
the Fermi -- Pasta -- Ulam (FPU) $\beta$ -- model from the point of
view of stability of its simplest periodic orbits (SPOs). In
particular, we show that there is a relatively high $q$ mode
$(q=2(N+1)/{3})$ of the linear lattice, having one particle fixed
every two oppositely moving ones (called SPO2 here), which can be
exactly continued to the nonlinear case for
$N=5+3m,\;m=0,1,2,\ldots$ and whose first destabilization, $E_{2u}$,
as the energy (or $\beta$) increases for {\it any} fixed $N$,
practically {\it coincides} with the onset of a ``weak'' form of
chaos preceding the break down of FPU recurrences, as predicted
recently in a similar study of the continuation of a very low
($q=3$) mode of the corresponding linear chain. This energy
threshold per particle behaves like $\frac{E_{2u}}{N}\propto
N^{-2}$. We also follow exactly the properties of another SPO (with
$q=(N+1)/{2}$) in which fixed and moving particles are interchanged
(called SPO1 here) and which destabilizes at higher energies than
SPO2, since $\frac{E_{1u}}{N}\propto N^{-1}$. We find that,
immediately after their first destabilization, these SPOs have
different (positive) Lyapunov spectra in their vicinity. However, as
the energy increases further (at fixed $N$), these spectra converge
to {\it the same} exponentially decreasing function, thus providing
strong evidence that the chaotic regions around SPO1 and SPO2 have
``merged'' and large scale chaos has spread throughout the lattice.
Since these results hold for $N$ arbitrarily large, they suggest a
direct approach by which one can use local stability analysis of
SPOs to estimate the energy threshold at which a transition to
ergodicity occurs and thermodynamic properties such as Kolmogorov --
Sinai entropies per particle can be computed for similar one --
dimensional lattices.
\end{abstract}

\section{Introduction}\label{intro}

The transition to widespread chaos in Hamiltonian systems of many
degrees of freedom has been the subject of intense investigation
for more than fifty years, see e.g.
\cite{Lichtenberg,MacKay_1987,Wiggins,Simo}. It received great
impetus following the pioneering work of Fermi, Pasta and Ulam
\cite{Fermi_1955}, who were the first to study thermalization in
one -- dimensional lattices of $N$ particles, with linear and
nonlinear nearest neighbor forces, as a parameter multiplying the
nonlinear terms in the equations of motion becomes greater than
zero. Surprisingly, they observed that when this parameter is
relatively small, energy equipartition does {\it not} occur even
after very long integration times, as only a small number of (low
$q$) modes of the corresponding linear lattice recurrently
exchange the total energy among them. Of course, when the
nonlinearity or the energy exceeds a certain threshold, these so
called FPU recurrences break down, large scale chaos prevails and
a type of ergodicity sets in, whereby almost every orbit explores
almost all of the available phase space of the system.

One of the first attempts to explain this phenomenon is due to
Izrailev and Chirikov \cite{Izrailev}, who argued that the breakdown
of FPU recurrences is related to the overlap of major resonances
known to lead to large scale chaos in $N$ -- degree of freedom
Hamiltonian systems \cite{Chirikov}. As it was later realized,
however, a weaker form of chaos caused by the interaction of the
first few FPU modes (with low $q$ in Fourier space) appears to be
sufficient for equipartition among all modes to occur
\cite{De_Luca_1995,De_Luca_2002}. Finally, very recently, Flach and
coworkers \cite{Flach_2005} discovered that this transition to the
so called ``weak'' chaos, in fact, coincides with the first
destabilization of one of the lowest ($q=3$) normal mode of the
linear lattice, as it is continued by increasing the nonlinearity
parameter. In fact, their results apply more generally to the lowest
$q\ll N$ modes which also turn out to be highly localized in $q$ --
space. These nonlinear modes represent examples of what we call
simple periodic orbits (SPOs), where all particles return to their
initial condition after only one oscillation, i.e. all their mutual
rotation numbers are unity \cite{Antonopoulos_2006}. Such lowest $q$
mode SPOs, were actually termed $q$ -- breathers due to their
exponential localization in Fourier space \cite{Flach_2005}.

In this paper, we investigate further the connection between local
and global dynamics of the FPU lattice, by studying the stability
properties of its SPOs. In particular, we consider the FPU
Hamiltonian

\begin{equation}\label{FPU_Hamiltonian}
H=\frac{1}{2}\sum_{j=1}^{N}\dot{x}_{j}^{2}+\sum_{j=0}^{N}\biggl
(\frac{1}{2}(x_{j+1}-x_{j})^2+\frac{1}{4}\beta(x_{j+1}-x_{j})^4\biggr)=E
\end{equation}
often called the FPU $\beta$ -- model, as it only contains the
term with quartic nonlinearities. The $x_{j}$ represents the
displacement of the $j$th particle from its equilibrium position,
$\dot{x}_{j}$ is the corresponding velocity, $\beta$ is a positive
real constant and $E$ is the total energy. As with the original
FPU problem, we will concern ourselves only with fixed boundary
conditions, whereby particles with index $j=0, N+1$ are stationary
for all time.

In particular, we will examine an SPO which keeps every {\it third}
particle fixed, while the two in between are performing exactly
opposite motions. This mode was studied in \cite{Bivins} where its
stability was analyzed by means of Mathieu equations and has also
been discussed in \cite{Kosevich} in connection with the occurrence
of sinusoidal waves in nonlinear lattices. This solution, called
SPO2 from here on, will be compared to an orbit we call SPO1, which
keeps every two particles fixed, with the ones in between executing
exactly opposite oscillations. This latter one was originally
mentioned in a paper by Ooyama et. al. \cite{Ooyama} and later
studied analytically by Budinsky and Bountis \cite{Budinsky} to
determine its stability properties in the thermodynamic limit of
large $N$ and $E$ with $\frac{E}{N}$ fixed. Recently, it was
revisited by Antonopoulos et al. \cite{Antonopoulos_2006}, in a
study of different SPOs and different Hamiltonians, from the
viewpoint of connecting their local and global stability properties.

Our first result about the SPO2 orbit is that the energy per
particle of its first destabilization goes to zero {\it faster} than
SPO1, by a law $\frac{E_{2u}}{N}\propto N^{-2}$ in contrast to the
SPO1 orbit whose law is $\frac{E_{1u}}{N}\propto N^{-1}$ as
$N\rightarrow\infty$ \cite{Antonopoulos_2006, Budinsky}. This
implies that if chaos is to spread in the nonlinear lattice as a
result of the destabilization of SPOs, it might be more useful to
look closely at the properties of SPO2, as that becomes unstable
much earlier than SPO1, as $N$ increases.

Remarkably enough, when we do this we discover that the energy (or
$\beta$) values of the first destabilization as a function of $N$
practically coincide with those found by \cite{De_Luca_1995,
De_Luca_2002} for the transition to ``weak'' chaos and
\cite{Flach_2005} for the destabilization of the $q=3$ mode. Our
numerical results and their analytical formula are in excellent
agreement.

We then examine the dynamics in more detail, following the first
destabilization of SPO1 and SPO2 at $E=E_{1u}$ and $E=E_{2u}$
respectively, at high enough $N$. In particular, we choose initial
conditions in the vicinity of these orbits and find that the
(positive) Lyapunov exponents, at energies just above $E_{1u}$ and
$E_{2u}$, fall off to zero following distinct curves, both for
SPO1, which destabilizes by a period -- doubling type of
bifurcation and SPO2, which exhibits a {\it complex} instability
with its monodromy eigenvalues exiting the unit circle in complex
conjugate pairs. However, as the energy increases further, the
Lyapunov spectra near SPO1 and SPO2 begin to {\it converge}, at
some $E>E_{1u}>E_{2u}$, to the same functional form, implying that
the chaotic regions of SPO1 and SPO2 have ``merged'' and large
scale chaos has spread in phase space.

The function to which the Lyapunov spectra converge is a nearly
exponentially decaying curve of the form
\begin{equation}\label{Lyap_Spec_convergence_exp_function}
L_{i}(N)\propto e^{-\alpha\frac{i}{N}}, i=1,2,\ldots,K(N)
\end{equation}
at least up to $K(N)\approx\frac{3N}{4}$, as we have discussed at
length in a recent publication \cite{Antonopoulos_2006}. This
function provides, in fact, an invariant of the dynamics, in the
sense that, in the thermodynamic limit, we can use it to evaluate
the average of the positive Lyapunov exponents (i.e. the
Kolmogorov -- Sinai entropy per particle) and find that it is a
constant characterized by the value of the exponent $\alpha$
appearing in (\ref{Lyap_Spec_convergence_exp_function}).

Thus, we argue that studying the local dynamics around some of the
simplest periodic orbits which destabilize at low energies, opens
a ``window'' into the ``global'' dynamics of nonlinear lattices.
Furthermore, by computing and comparing Lyapunov spectra in their
vicinity, it is possible to gain valuable insight into the
conditions for large scale chaos and ergodicity, so that we may be
able to define probability distributions and compute thermodynamic
properties of the lattice, as $E$ and $N$ increase indefinitely
with $E/N$ fixed.

Our paper is organized as follows: In section \ref{sec_1} we provide
analytical expressions of the SPO1 and SPO2 solutions under study
and describe in detail their stability properties for an arbitrarily
large number of particles of the FPU $\beta$ -- model with fixed
boundary conditions. Comparing with similar findings in the
literature, we observe that our results accurately predict the onset
of ``weak'' chaos preceding the break down of FPU recurrences,
although the reasons for this agreement are still under
investigation. In section \ref{sec_2}, we use our results on the
convergence of Lyapunov spectra, as the energy is increased beyond
the destabilization thresholds of SPO1 and SPO2 to estimate the
onset of large scale chaos and thermodynamic behavior in the lattice
and in section \ref{concl} we present our conclusions. We thus
believe that today, $50$ years after its famous discovery, the Fermi
-- Pasta -- Ulam problem and its transition from recurrences to
globally chaotic behavior is still very much alive as a topic of
intense research into some truly fundamental questions connecting
classical and statistical mechanics \cite{Berman}.

\section{Simple Periodic Orbits and Stability Analysis}\label{sec_1}

\subsection{Analytical results for SPO1}\label{subsec_1}

Let us start by describing briefly some analytical results
concerning SPO1, as this particular mode has been studied recently
by Antonopoulos et al. \cite{Antonopoulos_2006} and also previously
in \cite{Ooyama,Budinsky}.

We consider, for this reason, a one -- dimensional lattice of $N$
particles with equal masses and nearest neighbor interactions with
quartic nonlinearities ($\beta$ -- model) which is given by the FPU
Hamiltonian (\ref{FPU_Hamiltonian}), with fixed boundary conditions
\begin{equation}\label{FPU_fixed_boundary_conditions_SPO1}
x_{0}(t)=x_{N+1}(t)=0,\forall t.
\end{equation}

For $\beta=0$, Hamiltonian (\ref{FPU_Hamiltonian}) describes a
system of coupled harmonic oscillators and hence all solutions can
be written as combinations of $N$ independent normal modes whose
individual energies are constant in time. Since, in that case, the
spectrum has frequencies $\omega_q=2\sin(\pi q/2(N+1))$ which are
rationally independent, all solutions are quasiperiodic, and hence
the only strictly periodic solutions are the normal modes, with
frequencies $\omega_q,q=1,2,\ldots,N$
\cite{Izrailev,Flach_2005,Berman}. That these modes can be
continued for $\beta>0$ is a consequence of a famous theorem by
Lyapunov \cite{Lyapunov}, based on the assumption that no ratio of
linear frequencies $\omega_q/\omega_r$ is an integer, for
$q,r=1,2,\ldots,N$, which holds in this case. These solutions are
examples of what we call Simple Periodic Orbits (SPOs), in which
all particles return to their starting point after one maximum
(and one minimum) in their oscillation \cite{Antonopoulos_2006}.

Let us consider one such orbit -- we shall call SPO1 -- which is
specified by the conditions
\begin{equation}\label{FPU_non_lin_mode_fixed_boundary_conditions_SPO1}
\hat{x}_{2j}(t)=0,\;\hat{x}_{2j-1}(t)=-\hat{x}_{2j+1}(t)\equiv\hat{x}(t),\;j=1,\ldots,\frac{N-1}{2}
\end{equation}
and exists for all odd $N$, keeping every even particle stationary
at all times. It is not difficult to show that this is, in fact, the
$q=(N+1)/2$ mode of the linear lattice with frequency
$\omega_q=\sqrt{2}$. The remarkable property of this solution is
that it is continued in precisely the same spatial configuration in
the nonlinear lattice as well, due to the form of the equations of
motion associated with Hamiltonian (\ref{FPU_Hamiltonian}),
\begin{equation}\label{FPU_eq_motion_SPO1}
\ddot{x}_{j}(t)=x_{j+1}-2x_{j}+x_{j-1}+\beta\Bigl((x_{j+1}-x_{j})^3-(x_{j}-x_{j-1})^3\Bigr),\;j=1,\ldots,N
\end{equation}
which reduce, upon using
(\ref{FPU_non_lin_mode_fixed_boundary_conditions_SPO1}) with
(\ref{FPU_fixed_boundary_conditions_SPO1}) to a single second order
nonlinear differential equation for $\hat{x}(t)$,
\begin{equation}\label{FPU_single_equation_SPO1}
\ddot{\hat{x}}(t)=-2\hat{x}(t)-2\beta\hat{x}^{3}(t)
\end{equation}
describing the oscillations of all moving particles of SPO1, with
$j=1,3,5,\ldots,N$. For the stationary particles
$j=2,4,6,\ldots,N-1$ of course, we have $\hat{x}(t)=0,\forall
t\ge0$. The solution of (\ref{FPU_single_equation_SPO1}) is well
known in terms of Jacobi elliptic functions \cite{Abramowitz} and
can be written as
\begin{equation}\label{sol_FPU_single_equation_SPO1}
\hat{x}(t)=\mathcal{C}\cn(\lambda t,{\kappa}^{2})
\end{equation}
where
\begin{equation}\label{FPU_C_and_lambda_SPO1}
\mathcal{C}^{2}=\frac{2{\kappa}^{2}}{\beta(1-2{\kappa}^{2})},\
\lambda^{2}= \frac{2}{1-2{\kappa}^{2}}
\end{equation}
and ${\kappa}^{2}$ is the modulus of the $\cn$ elliptic function.
The energy per particle of SPO1 is then found to be
\begin{equation}\label{FPU_energy_per_perticle_SPO1}
\frac{E}{N+1}=\frac{1}{4}\mathcal{C}^{2}(2+\mathcal{C}^{2}\beta)=\frac{\kappa^{2}(1-\kappa^{2})}{\beta(1-2{\kappa}^{2})^{2}}
\end{equation}
by substituting simply the solution $\hat{x}(t)$ of
(\ref{sol_FPU_single_equation_SPO1}) in Hamiltonian
(\ref{FPU_Hamiltonian}).

\begin{figure}[ht]
\begin{center}
\includegraphics{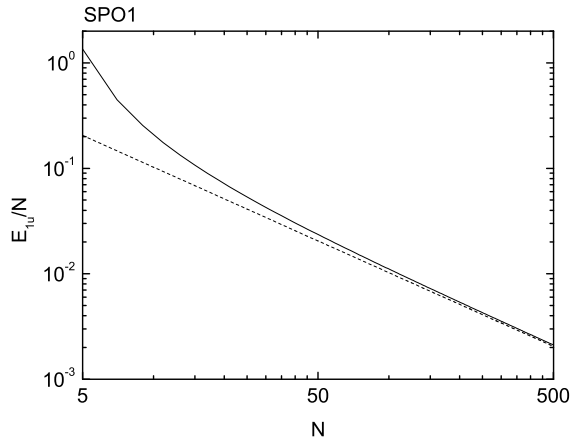} \vspace{4.8cm}
\end{center}
\caption{The solid curve is the energy per particle
$\frac{E_{1u}}{N}$ of the first destabilization of the SPO1 for
$\beta=1$ while the dashed line is the function
$\propto\frac{1}{N}$.}\label{graph_1}
\end{figure}

The linear stability analysis of the SPO1 mode is straightforward
and was carried out recently in \cite{Antonopoulos_2006} using
Lam\'{e} equations, Hill's determinants and Floquet theory. Plotting
the first destabilization energy for this orbit as a function of $N$
with solid lines in Fig. \ref{graph_1}, we observe that the
corresponding energy density threshold $\frac{E_{1u}}{N}$ decreases
with $N$ following a simple power law $\propto 1/N$ (dashed line).
Following such an approach, we find, for example, for $\beta=1$ and
$N=11$, that SPO1 destabilizes for the first time when
$E_{1u}\approx1.93$.

\subsection{The solution of SPO2}\label{subsec_2}

Let us now turn to the second simple periodic orbit studied in
this paper, which we call SPO2. We impose again fixed boundary
conditions and consider lattices consisting of
$N=5+3m,\;m=0,1,2,\ldots$ particles, in which every third particle
is fixed, while the two in between move in opposite directions as
follows
\begin{eqnarray}
x_{3j}(t)&=&0,\;j=1,2,3\ldots,\frac{N-2}{3},\label{FPU_FBC_AB_mode_1}\\
x_{j}(t)&=&-x_{j+1}(t)=\hat{x}(t),\;j=1,4,7,\ldots,N-1.\label{FPU_FBC_AB_mode_2}
\end{eqnarray}
This solution, in fact, corresponds to the $q=2(N+1)/3$ normal
mode with frequency $\omega_q=\sqrt{3}$ of the linear system and
can also be continued in exactly the same form in the nonlinear
case $\beta>0$, due to the symmetry of the equations of motion,
\begin{equation}\label{FPU_eq_motion_SPO_2_FBC}
\ddot{x}_{j}(t)=x_{j+1}-2x_{j}+x_{j-1}+\beta\Bigl((x_{j+1}-x_{j})^3-(x_{j}-x_{j-1})^3\Bigr),\;j=1,\ldots,N
\end{equation}
which, under the above conditions, (\ref{FPU_FBC_AB_mode_1}),
(\ref{FPU_FBC_AB_mode_2}), collapse to a single second order
nonlinear differential equation very similar to
(\ref{FPU_single_equation_SPO1})
\begin{equation}\label{FPU_single_equation_SPO 2_FBC}
\ddot{\hat{x}}(t)=-3\hat{x}(t)-9\beta\hat{x}^{3}(t).
\end{equation}

As before, this equation describes the moving particles of the
lattice, while the stationary ones satisfy  $\hat{x}(t)=0,\forall
t\ge0$, for $j=3,6,9,\ldots,N-2$. The solution of equation
(\ref{FPU_single_equation_SPO 2_FBC}) is again given in terms of the
Jacobi elliptic functions \cite{Abramowitz} and is written in the
form
\begin{equation}\label{sol_FPU_single_equation_SPO 2_FBC}
\hat{x}(t)=\mathcal{C}\cn(\lambda t,{\kappa}^{2})
\end{equation}
where
\begin{equation}\label{FPU_C_and_lambda_SPO 2_FBC}
\mathcal{C}^{2}=\frac{2{\kappa}^{2}}{3\beta(1-2{\kappa}^{2})},\
\lambda^{2}= \frac{3}{1-2{\kappa}^{2}}
\end{equation}
and ${\kappa}^{2}$ is, again, the modulus of the $\cn$ elliptic
function. The energy per particle of the SPO2 mode is found to be
now
\begin{equation}\label{FPU_energy_per_perticle_SPO 2_FBC}
\frac{E}{N+1}=\frac{2\kappa^{2}(1-{\kappa}^{2})}{3\beta(1-2{\kappa}^{2})^{2}}
\end{equation}
by simply substituting the solution $\hat{x}(t)$ of
(\ref{FPU_single_equation_SPO 2_FBC}) in Hamiltonian
(\ref{FPU_Hamiltonian}).

In order to perform the linear stability analysis of the SPO2 mode
we set $x_{j}=\hat{x}_{j}+y_{j}$ in the equations of motion
(\ref{FPU_eq_motion_SPO_2_FBC}) and keep up to linear terms in the
small displacement variable $y_{j}$. We thus get the variational
equations for this orbit in the form
\begin{eqnarray}
\ddot{y}_j(t)&=& A_3 y_{j-1}+A_1y_j+A_2y_{j+1},\;j=1,4,7,\ldots,N-1,\label{FPU_variational_equations_SPO 2_FBC_1}\\
\ddot{y}_j(t)&=& A_2 y_{j-1}+A_1 y_j+A_3 y_{j+1},\;j=2,5,8,\ldots,N,\label{FPU_variational_equations_SPO 2_FBC_2}\\
\ddot{y}_j(t)&=& A_3 (y_{j-1}-2 y_j+
y_{j+1}),\;j=3,6,9,\ldots,N-2\label{FPU_variational_equations_SPO
2_FBC_3}
\end{eqnarray}
where $y_{0}=y_{N+1}=0$ and
\begin{eqnarray}
A_1&=&-2-15\beta\hat{x}^2(t),\\
A_2&=&1+12\beta\hat{x}^2(t),\\
A_3&=&1+3\beta\hat{x}^2(t).
\end{eqnarray}

Unfortunately, it is not as easy to uncouple the above linear system
of differential equations and study the stability of the SPO2, in
terms of independent Lam\'{e} equations, as we were able to do with
SPO1 \cite{Antonopoulos_2006}. We can, however, compute numerically
with arbitrary accuracy and for every given
$N=5+3m,\;m=0,1,2,\ldots$ the complex eigenvalues
$\lambda_{i},\;i=1,\ldots,2N$ of the corresponding monodromy matrix
and characterize the stability of the SPO2 by their position on the
complex plane with regard to the unit circle.

\begin{figure}[ht]
\begin{center}
\includegraphics{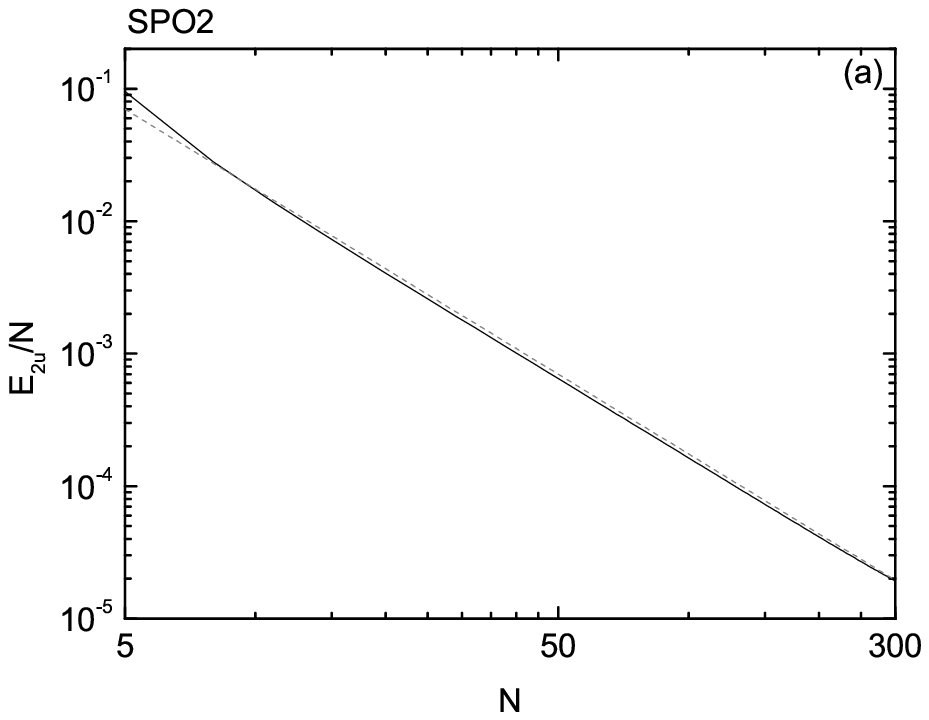} \includegraphics{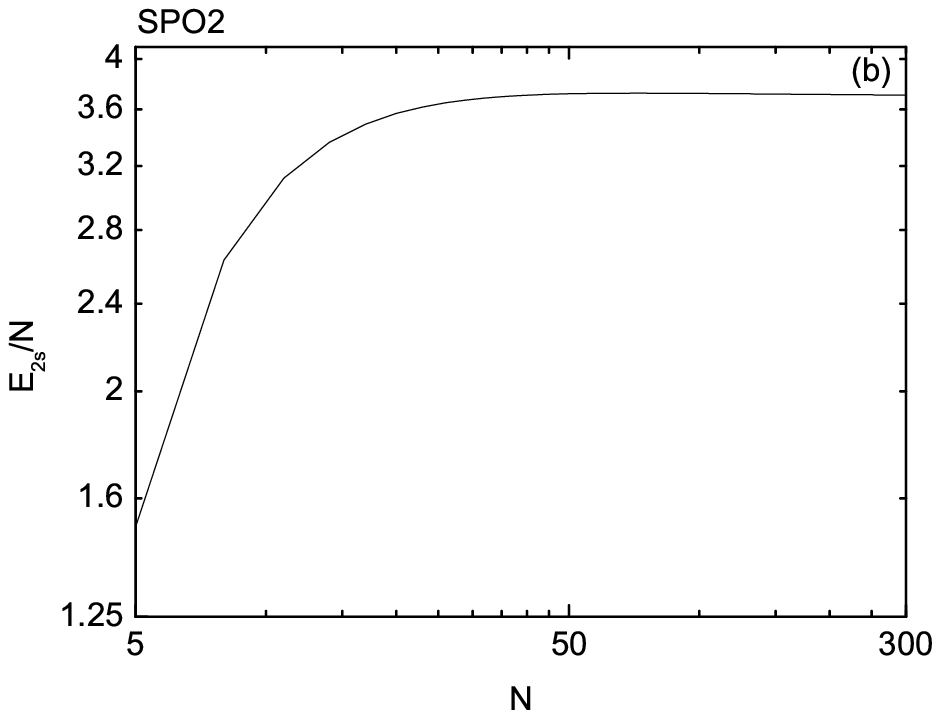} \vspace{4.8cm}
\end{center}
\caption{(a) The solid curve is the energy density
$\frac{E_{2u}}{N}$ of the first destabilization of the SPO2 obtained
by the numerical evaluation of the eigenvalues of the monodromy
matrix, while the dashed line is $\propto\frac{1}{N^{2}}$. (b) The
energy density $\frac{E_{2s}}{N}$ of the first restabilization of
the SPO2 obtained by the same method as in (a). In this figure
$\beta =1$.} \label{graph_2}
\end{figure}

We have thus computed, for many values of
$N=5+3m,\;m=0,1,2,\ldots,98$, the energy $E_{2u}(N)$ of the first
destabilization of the SPO2 for $\beta=1$ and have plotted the
results with solid lines in Fig. \ref{graph_2}(a). As we see, the
energy density $\frac{E_{2u}}{N}$ at the first instability decreases
following a power -- law $\propto 1/N^{2}$ (dashed line) which is
faster than the SPO1 solution we discussed earlier, see Fig.
\ref{graph_1}. Following this approach, we find, for $\beta=1$ and
$N=11$, that SPO2 destabilizes for the first time when
$E_{2u}\approx0.153$.

Interestingly enough, if we calculate the eigenvalues of the
monodromy matrix of the SPO2 for greater energies, we find that it
becomes again stable, beyond a new critical energy $E_{2s}(N)$. In
Fig. \ref{graph_2}(b) we plot this restabilization energy density of
SPO2 as a function of $\log N$ and observe that it approaches a
constant as $N$ tends to infinity.

Pursuing further this restabilization phenomenon, we have estimated
the ``size'' of the islands of stability around SPO2 for energies
above the energy threshold $E_{2s}(N)$, using the method of the
Smaller ALignment Index (SALI)
\cite{Skokos_2001,Skokos_1_2003,Skokos_2_2003,Skokos_2004}. This
index has proved to be very efficient for distinguishing rapidly and
with certainty regular vs. chaotic orbits, as it exhibits completely
different behavior in these two cases: It fluctuates around non --
zero values for regular orbits, while it converges exponentially to
zero for chaotic orbits. SALI is particularly useful in the case of
many degrees of freedom, where very few such methods are available
beyond the cumbersome and often inconclusive calculation of the
maximal Lyapunov exponent.

We thus observe the following: As the energy $E$ increases beyond
the restabilization threshold $E_{2s}(N)$, for fixed $N$, the
``size'' of the island around SPO2 changes very little, compared
with the growth of the system's available phase space. Moreover, if
we keep $\frac{E}{N}$ fixed, thus holding the ``radius'' of the
energy surface nearly constant, we find that the ``radius'' of the
SPO2 island diminishes as a function of $E$. This is done by
changing one of the particles' position and momentum by $\Delta x$
and $\Delta p_x$ away from its SPO2 values while keeping the energy
constant and using SALI to estimate $\Delta x_{max}$ at which we
reach chaos (e.g. with $\frac{E}{N}=4$ and $N=5,8,11,14$, we find
respectively $\Delta x_{max}\approx0.01,0.005,0.0025,0.002$). We,
therefore, conclude that the island of ordered motion around the
SPO2 solution should be of no consequence to the statistical
properties of the lattice, such as ergodicity and the definition of
thermodynamic quantities.

\begin{figure}[]
\begin{center}
\includegraphics{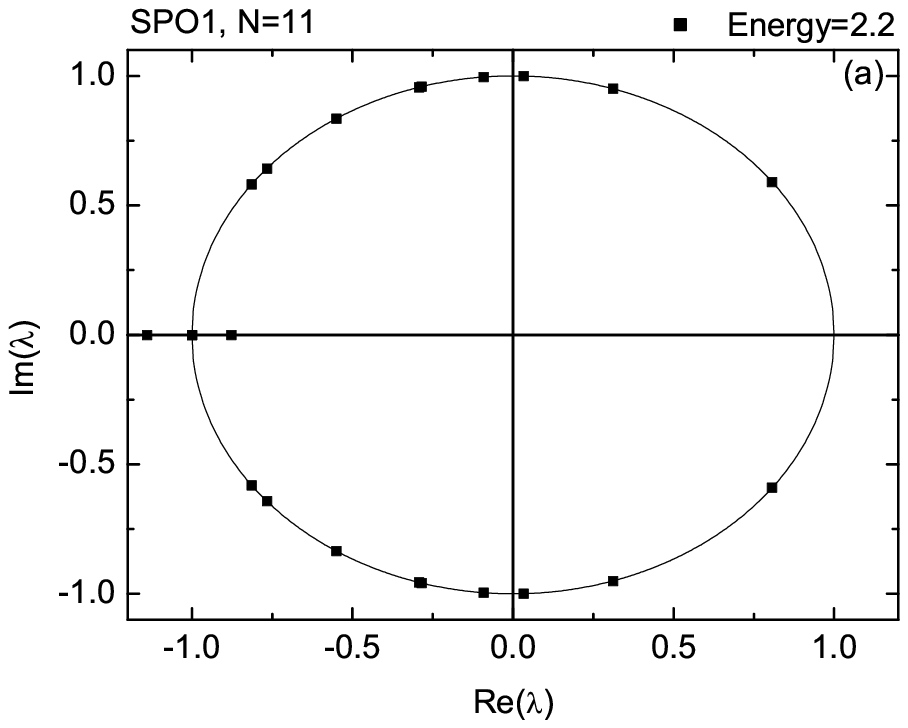} \includegraphics{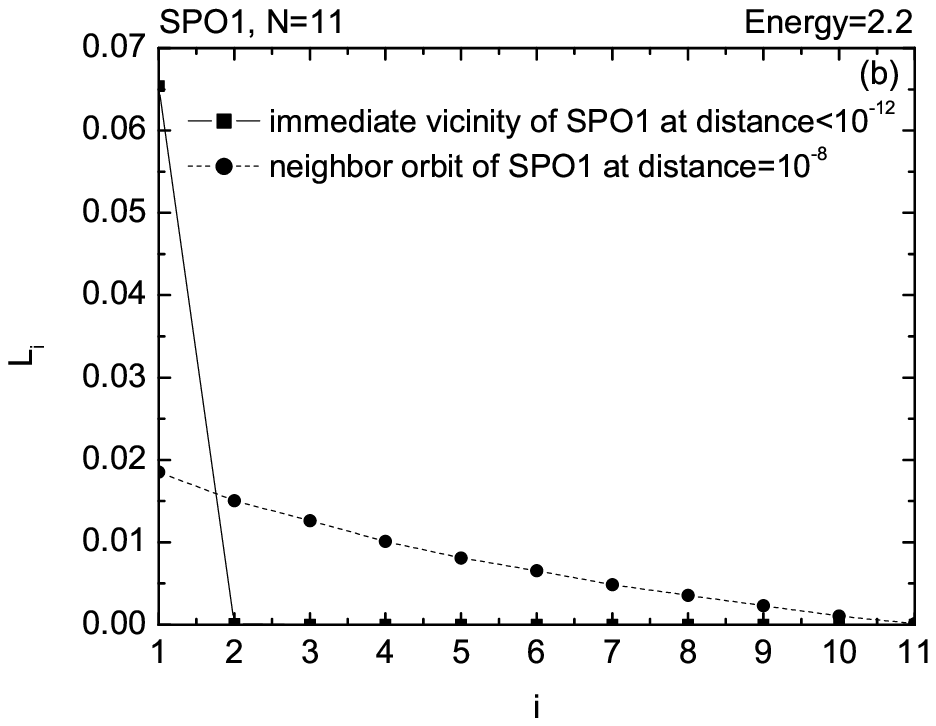} \includegraphics{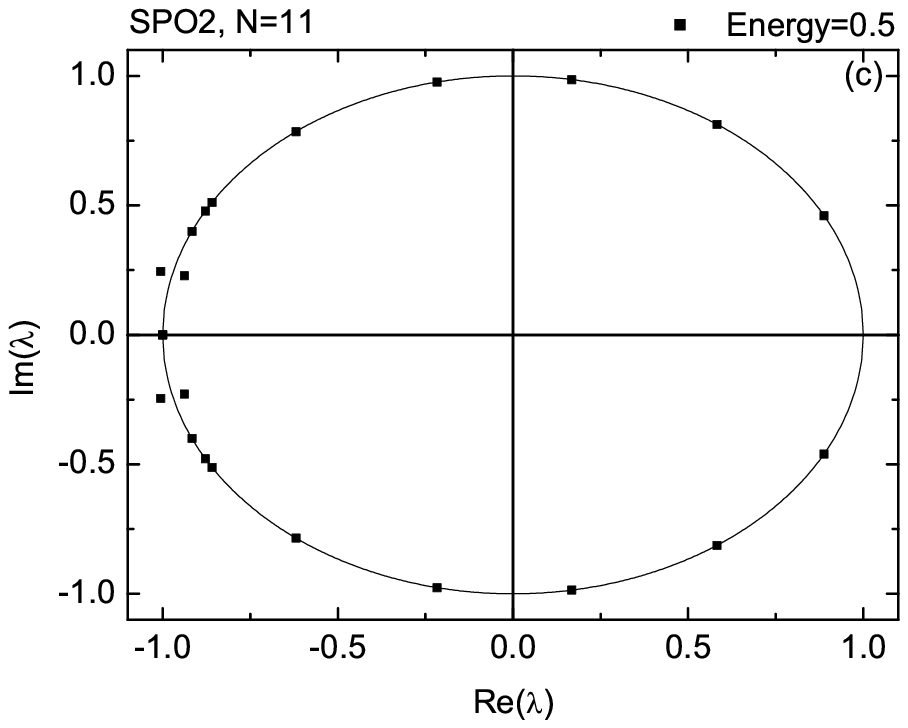} \includegraphics{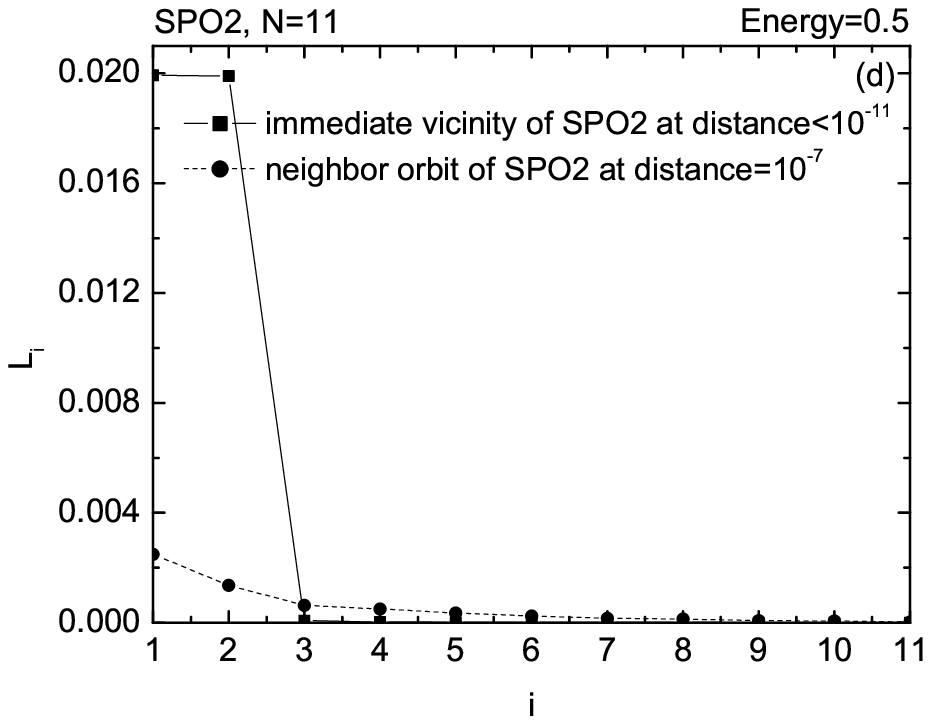} \vspace{11cm}
\end{center}
\caption{(a) The period -- doubling bifurcation of SPO1 for $N=11$
at the energy $E=2.2$. (b) The Lyapunov spectrum of two orbits, one
starting very close to SPO1 and another a little further away, for
the same $N$ and $E$ as in panel (a). (c) The complex instability of
SPO2 for $N=11$ at the energy $E=0.5$ after its first
destabilization. (d) The Lyapunov spectrum of two similar orbits,
one in the immediate vicinity and another more distant from SPO2 for
the same $N$ and $E$ as in panel (c). In this figure $\beta =1$.
}\label{graph_3}
\end{figure}

Moreover, as we observe in Fig. \ref{graph_3}(a), (c), the kinds of
bifurcation leading to instability for these SPOs are very
different: In the case of SPO1, Fig. \ref{graph_3}(a) shows that the
bifurcation is of the period -- doubling type, as one pair of real
eigenvalues is seen to exit the unit circle at -1, while Fig.
\ref{graph_3}(c) shows that we have complex instability in the case
of SPO2. This is also indicated by the positive Lyapunov exponents
in the {\it immediate} vicinity of the SPOs (about $10^{-12}$ from
them), which are closely connected with the eigenvalues of the
corresponding variational equations and are shown in Fig.
\ref{graph_3}(b), (d) with solid lines. Of course, moving away from
the two modes (within a range of about $10^{-11}$ -- $10^{-2}$), the
Lyapunov spectra change into the familiar form of two smoothly
decaying, evidently different curves, plotted with dashed lines in
Fig. \ref{graph_3}(b), (d). This fact suggests that the chaotic
regions near these modes are separated from each other in phase
space.

\begin{figure}[]
\begin{center}
\includegraphics{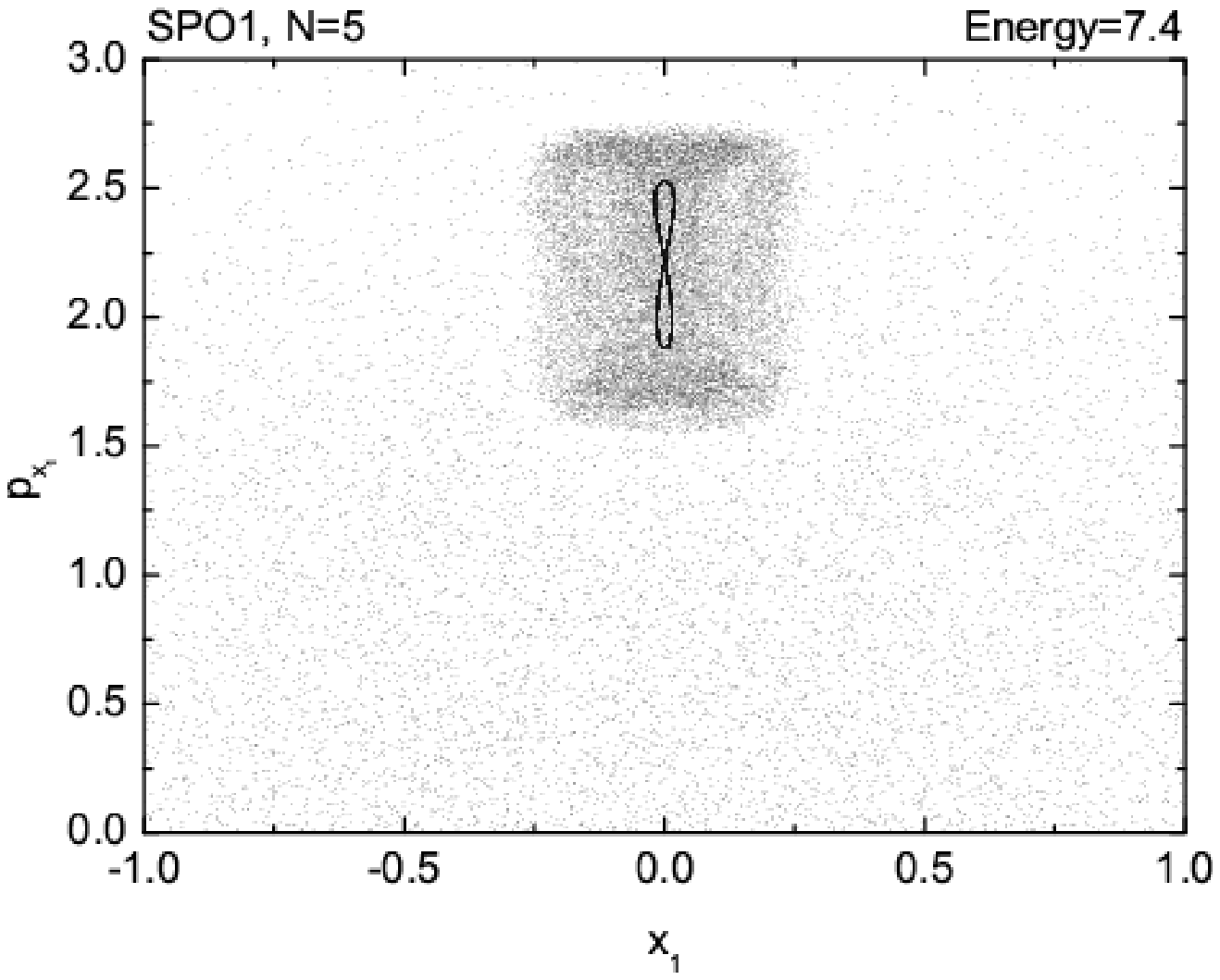} \vspace{4.8cm}
\end{center}
\caption{The ``figure eight'' chaotic region for initial conditions
in the immediate vicinity of SPO1 ($\simeq10^{-5}$), a vague
resemblance to ``figure eight'' for initial conditions a little
further away ($\simeq10^{-1}$) and a large scale chaotic region in
the energy surface for initial conditions more distant ($\simeq1$)
for $N=5$ particles, when it is unstable, on the Poincar\'{e}
surface of section ($x_1,\dot{x}_1$) computed at times when $x_3=0$.
In this picture we integrated our orbits up to $t_n=10^5$ in the
energy surface $E=7.4$.}\label{graph_4}
\end{figure}

It is quite interesting to observe that the chaotic region about an
unstable SPO can be isolated in phase space from the chaotic motion
occurring in different domains. In fact, there may be several such
domains embedded into each other. For example, in the $N=5$ FPU
$\beta$ -- model, when SPO1 becomes unstable, a ``figure eight''
chaotic region becomes clearly visible in its immediate
neighborhood, on a Poincar\'{e} surface of section ($x_1,\dot{x}_1$)
taken at times when $x_3=0$ (see Fig. \ref{graph_4}). Even though
the SPO1 mode is unstable, nearby orbits oscillate about it for very
long times, forming eventually the ``figure eight'' we see in the
picture.

More surprisingly however, starting at points a little further away,
a different chaotic domain is observed which bears a vague
resemblance to the ``figure eight'' and does not spread to the full
energy surface. Of course, if one chooses more distant initial
conditions a large scale chaotic region becomes evident on the
surface of section of Fig. \ref{graph_4}.

We have checked that the Lyapunov exponents in these regions are
quite different from each other, at least when one integrates the
equations of motion up to $t=10^5$. Of course, if an orbit lies on
the ``boundary'' between two of these domains, if integrated long
enough, it may drift from the inner to the outer chaotic region,
where its Lyapunov exponents are expected to change accordingly.

In section \ref{sec_2}, we will study in more detail the relative
location of chaotic domains in phase space and argue that these will
``overlap'' when their respective Lyapunov spectra begin to converge
as a function of increasing energy for fixed $N$.

\subsection{Comparison with results in the literature}\label{subsec_3}

It was shown very recently in \cite{Flach_2005} that the linear
modes of the FPU $\beta$ -- model can be continued as SPOs of the
corresponding  nonlinear lattice, i.e. as exact, time periodic
solutions, having nearly the same spatial configuration and
frequency as in the linear case. These new solutions are
characterized by exponential localization in the $q$ -- space of
the normal modes $Q_j(t),j=1,\ldots,2N$ and preserve their
stability for small enough $\beta>0$. In fact, the energy
threshold for the destabilization of the $q=3$ solution found by
\cite{Flach_2005} coincides with the ``weak'' chaos threshold
determined by de Luca and Lichtenberg in \cite{De_Luca_1995}.

In this section, we show that the energy threshold found in
\cite{De_Luca_1995} and \cite{Flach_2005} also appears to coincide
with the instability threshold of the SPO2 mode. This is somewhat
surprising since, in all studies of the breakdown of FPU
recurrences so far, the wave number $q$ of the periodic solutions
responsible for the transition to ``weak'' chaos is low ($q\leq
4$), while our SPO2 mode has a considerably higher wave number,
$q=2(N+1)/3$.

Thus, using different approaches, the authors of
\cite{De_Luca_1995} and \cite{Flach_2005} report an approximate
formula, valid to order $O(\frac{1}{N^2})$, for the
destabilization energy of the $q$ -- breather solution with wave
number $q=3$ given by

\begin{equation}\label{approximate_energy_unstable formula_flach_et_al}
E_c\approx\frac{\pi^2}{6\beta(N+1)}.
\end{equation}

\begin{figure}[ht]
\begin{center}
\includegraphics{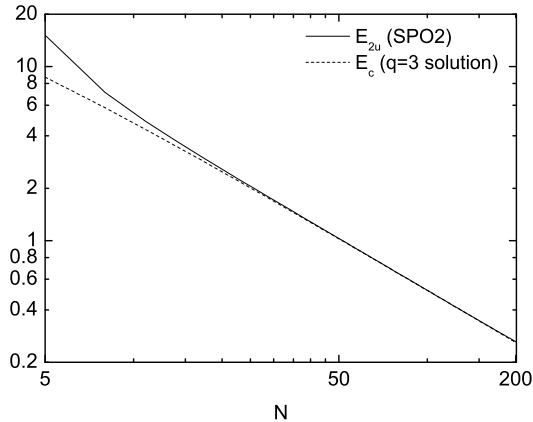} \vspace{4.8cm}
\end{center}
\caption{The solid curve corresponds to the energy $E_{2u}(N)$ of
the first destabilization of the SPO2 for $\beta=0.0315$ obtained by
the numerical evaluation of the eigenvalues of the monodromy matrix,
while the dashed line corresponds to the approximate formula
(\ref{approximate_energy_unstable formula_flach_et_al}) for the $q$
-- breather solution.}\label{graph_5}
\end{figure}

In Fig.\ref{graph_5} we compare the approximate formula
(\ref{approximate_energy_unstable formula_flach_et_al}) (dashed
line) with our destabilization threshold for SPO2 obtained by the
monodromy matrix analysis of subsection \ref{subsec_2} (solid line),
for $\beta=0.0315$, following \cite{Flach_2005}. We clearly observe
very good agreement between our numerical results and those of
(\ref{approximate_energy_unstable formula_flach_et_al}).

The reasons for this agreement are not yet clear to us. Of course,
if one wanted to look for a chaotic transition via the
destabilization of SPOs, it would be natural to analyze first SPO2
rather than SPO1, since SPO2 becomes unstable for lower energies.
Perhaps the power law $\frac{E_{2u}}{N}\propto 1/N^{2}$ is important
as it is the same in formula (\ref{approximate_energy_unstable
formula_flach_et_al}) as well as our SPO2 stability results. Still,
the agreement between the two curves in Fig. \ref{graph_5} cannot be
due only to the coincidence of power laws. The proportionality
factors in the corresponding formulas must also be nearly the same.

\section{Convergence of Lyapunov spectra}\label{sec_2}

Let us now start from the neighborhood of the SPO1 and SPO2 modes
and examine systematically the onset of large scale chaos in the
system as the energy is increased above the instability thresholds
$E_{1u}(N)$ and $E_{2u}(N)$. To do this, we need to evaluate the
Lyapunov exponents $L_i,\;i=1,\ldots,2N$ near the SPOs (ordered as
$L_1>L_2>\ldots>L_{2N}$), which measure the rate of exponential
divergence of nearby orbits in different directions of phase space
as time goes to infinity
\cite{Lichtenberg,Benettin_1980_1,Benettin_1980_2}.

Note that Fig. \ref{graph_3} reveals that when one calculates
Lyapunov exponents starting very close ($<10^{-12}$) to a periodic
orbit, one finds that they are closely related to the eigenvalues of
the monodromy matrix of the local dynamics. In Fig.
\ref{graph_3}(c), (d), we have plotted these quantities for the SPO2
mode at an energy where it has undergone a complex bifurcation and
has two pairs of eigenvalues off the unit circle on the complex
plane.

For comparison, we have also calculated in Fig. \ref{graph_3}(a),
the eigenvalues of the monodromy matrix of the SPO1 mode at an
energy where it is unstable with only one pair of eigenvalues off
the unit circle on the real negative axis. As we see in Fig.
\ref{graph_3}(b), very near this orbit the Lyapunov spectrum is
again in close agreement with the local results. Of course, in both
cases, starting with initial conditions a little further away yields
the true spectrum, as a smoothly decaying curve, which is an
invariant of the dynamics in that region (see the dashed lines in
Fig. \ref{graph_3}(b), (d)).

Thus, in the neighborhood of unstable SPOs, one can easily find
evidence of ``small'' scale chaos, which is visible at energies
where these orbits have just destabilized. This, however, is only a
local effect, which may have nothing to do with the chaotic behavior
anywhere else in the system. How could we use the dynamics near
SPOs, to obtain more global properties of the motion, like e.g. the
onset of large scale chaos in phase space?

One way to answer might be to test whether the chaotic regions of
the two SPOs become ``connected'' in phase space, above a certain
value of the energy. Evidence that such ``merging'' of chaotic
regions in phase space indeed occurs can be provided by their
maximal ($L_1$) Lyapunov exponents becoming equal and, more
specifically, by the convergence of the corresponding Lyapunov
spectra in their vicinity to an exponential function with a
characteristic exponent.

\begin{figure}[]
\begin{center}
\includegraphics{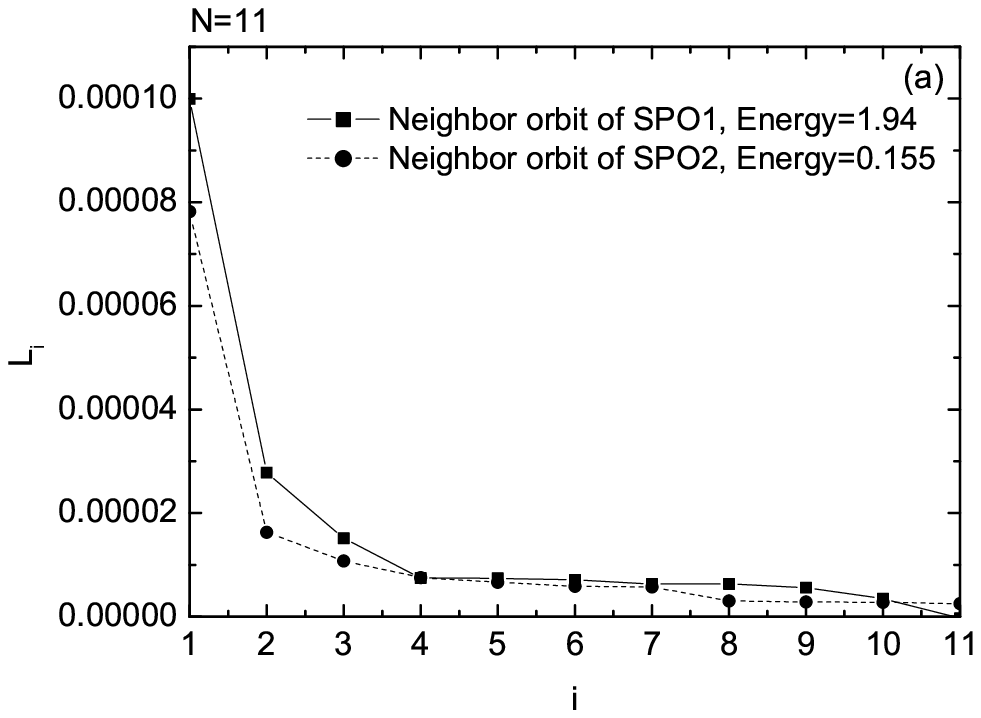} \includegraphics{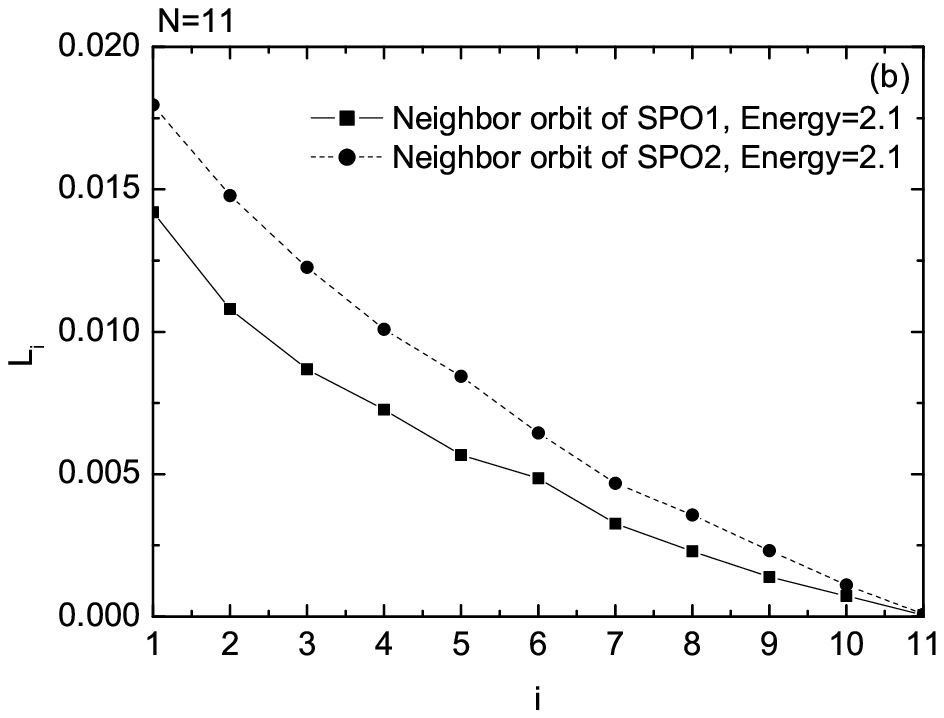} \includegraphics{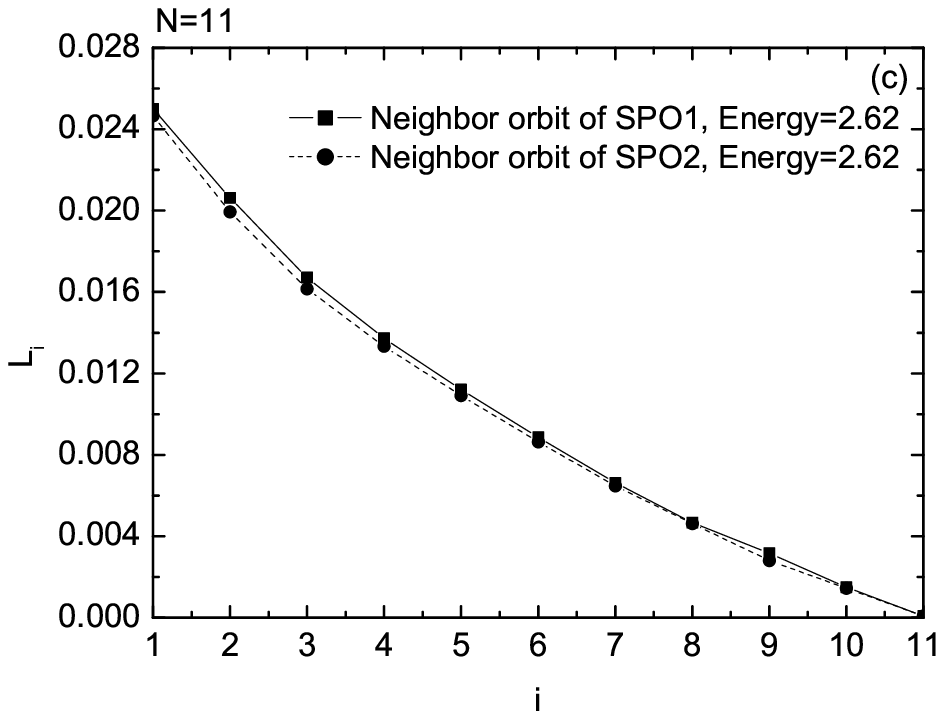} \includegraphics{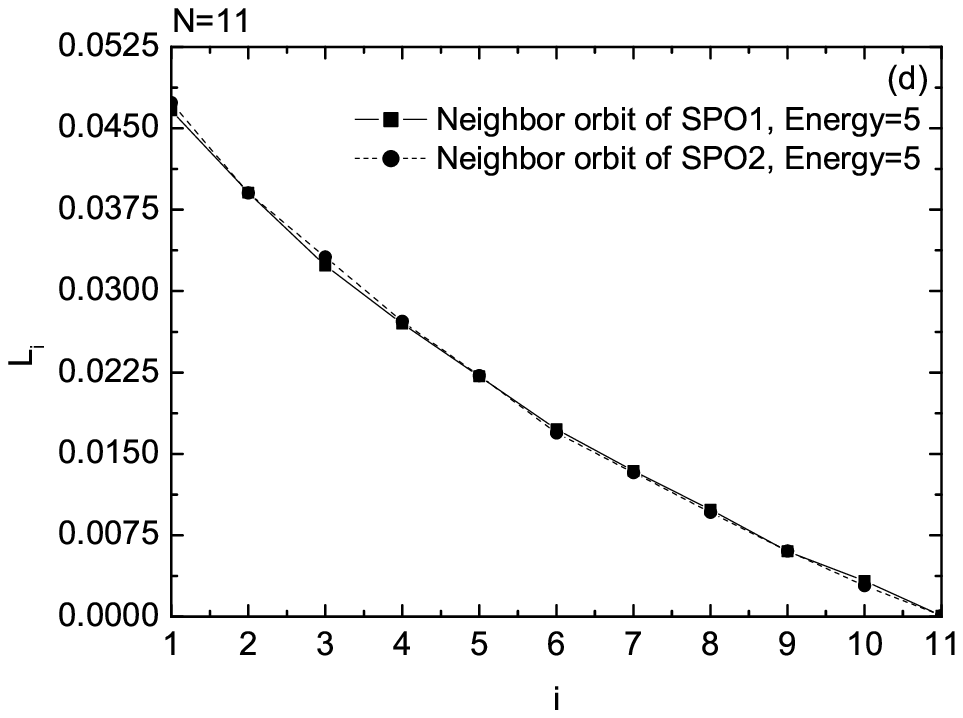} \vspace{11cm}
\end{center}
\caption{(a) Lyapunov spectra of neighboring orbits of SPO1 and SPO2
respectively for $N=11$ at energies $E=1.94$ and $E=0.155$, where
they respectively have just destabilized. (b) Same as in panel (a)
at energy $E=2.1$ for both SPOs, where the spectra are still
distinct. (c) Convergence of the Lyapunov spectra of neighboring
orbits of the two SPOs at energy $E=2.62$ where both of them are
unstable. (d) Coincidence of Lyapunov spectra continues at higher
energy $E=4$. All initial distances between nearby trajectories are
$\simeq10^{-2}$.}\label{graph_6}
\end{figure}

To see this, let us proceed to exhibit in Fig. \ref{graph_6}(a) the
Lyapunov spectra of two neighboring orbits of the SPO1 and SPO2
modes (all orbits in this figure start at distances $\simeq10^{-2}$
from the SPOs in phase space), for $N=11$ degrees of freedom and
energy values $E_{1}=1.94$ and $E_{2}=0.155$ respectively, where the
SPOs have just destabilized. As expected, in this case, the maximum
Lyapunov exponents $L_1$, are very small $(\approx10^{-4})$ and the
corresponding Lyapunov spectra are quite distinct.

Turning now to Fig. \ref{graph_6}(b), we observe that at the energy
value $E=2.1$, the Lyapunov spectra for both SPOs are much closer to
each other, even though their maximal Lyapunov exponents $L_1$ are
still different. Furthermore, in Fig. \ref{graph_6}(c), at $E=2.62$,
we see that the two spectra have nearly converged to the same
exponentially decreasing function,
\begin{equation}\label{Lyap_Spec_convergence_exp_function_again_1}
L_{i}(N)\propto e^{-\alpha\frac{i}{N}}, i=1,2,\ldots,K(N)
\end{equation}
at least up to $K(N)\approx\frac{3N}{4}$, while their maximal
Lyapunov exponents are virtually the same. The $\alpha$ exponents of
(\ref{Lyap_Spec_convergence_exp_function_again_1}) for the SPO1 and
SPO2 are found to be approximately $2.3$ and $2.32$ respectively.
Finally, Fig.\ref{graph_6}(d) shows that this coincidence of
Lyapunov spectra persists at higher energies.

\newpage

We regard this coincidence as an indication that the chaotic
regions of the two SPOs have ``merged'' in phase space in the
sense that they ``communicate'', as orbits starting initially in
the vicinity of one SPO may now visit the chaotic region of the
other. This is strong evidence of the existence of large scale
chaos in the FPU lattice, at least over the part of phase space
travelled by the SPO1 and SPO2 orbits, during their time
evolution.

\section{Conclusions}\label{concl}

In this paper we investigated the connection between local and
global dynamics in the FPU $\beta$ -- model from the point of view
of stability of its SPOs. Initially, we showed that a relatively
high $q$ mode of the linear lattice, with one particle fixed every
two oppositely moving ones, called SPO2, is stable for low energies
until it undergoes complex instability. In parallel, we also studied
the properties of another mode called SPO1, which keeps every two
particles fixed, with the ones in between executing exactly opposite
oscillations.

Our first result concerning these orbits is that the energy
threshold of the first destabilization of SPO2 goes to zero faster
than that of SPO1. Additionally, we discovered that, as a function
of $N$, the SPO2 destabilization threshold coincides with the one
found by other researchers for the transition to ``weak'' chaos and
the destabilization of the $q=3$ mode. This implies that if chaos is
to spread as a result of the destabilization of SPOs in the FPU
lattice, one might as well look closely at the properties of the
SPO2, as that becomes unstable much earlier than SPO1, as $N$
increases arbitrarily.

In order to examine their local dynamics in more detail, we raised
the energy above the destabilization of SPO1 and SPO2 and
calculated the Lyapunov spectra in their neighborhood. We thus
found that, as $E$ increases, the Lyapunov spectra in the
neighborhood of these SPOs appear to converge, at some relatively
low energy value, to the same functional form, implying that their
chaotic regions have ``merged'' and large scale chaos has spread
in the FPU lattice.

We, therefore, argue that by studying local dynamics near some of
the simplest periodic orbits which destabilize at low energies,
one can gain a better view of the ``global'' dynamics of nonlinear
lattices. Furthermore, by computing and comparing Lyapunov spectra
in the vicinity of such orbits, it is possible to obtain valuable
insight into the conditions for (or obstructions to) full scale
chaos and ergodicity, so that we may be able to define probability
distributions and compute thermodynamic properties of the lattice,
as the energy and the number of degrees of freedom increase
indefinitely, while the energy density is kept fixed.

\section{Acknowledgements}

This work was partially supported by the European Social Fund (ESF),
Operational Program for Educational and Vocational Training II
(EPEAEK II) and particularly the Program HERAKLEITOS, providing a
Ph.~D scholarship for one of us (C.~A.). C.~A. also acknowledges
with gratitude the $3$ month hospitality, March -- June $2005$, of
the ``Center for Nonlinear Phenomena and Complex Systems'' of the
University of Brussels where, also, in its facilities (ANIC4
computer cluster), the main part of the computer programs of this
work were executed. The second author (T.~B.) wishes to express his
gratitude to the Max Planck Institute of the Physics of Complex
Systems at Dresden, for its hospitality during his $3$ month visit
March -- June $2005$, when this work was initiated.


\end{document}